\documentclass[twocolumn,showpacs,floatfix]{revtex4} [10/03/2009]

\usepackage{amsmath,amssymb,amstext}
\usepackage{textcomp}
\usepackage{gensymb}
\usepackage{marvosym}
\usepackage{graphicx}
\usepackage{color}

\begin{document}

\title{\textbf{Quantum confinement effects in Si/Ge heterostructures with spatially ordered arrays of self-assembled quantum dots}}

\author{Oleksiy B. Agafonov$^1$, Christian Dais$^2$, Detlev Gr\"{u}tzmacher$^3$, and Rolf J. Haug$^1$}

\affiliation{$^1$Institut f\"{u}r Festk\"{o}rperphysik, Leibniz Universit\"{a}t Hannover, Appelstra\ss e~2, D-30167 Hannover, Germany}

\affiliation{$^2$Laboratory for Micro- and Nanotechnology, Paul Scherrer Institut, Villigen-PSI CH-5232, Switzerland}

\affiliation{$^3$Institute for Semiconductor Nanoelectronics, IBN-1, Forschungszentrum J\"{u}lich, D-52425 J\"{u}lich, Germany and J\"{u}lich Aachen Research Alliance on Fundamentals of Future Information Technology (JARAFIT)}
\date{\today}

\begin{abstract}

 Magnetotunneling spectroscopy was employed to probe the confinement in vertical Si/Ge double-barrier resonant tunneling diodes with regularly distributed Ge quantum dots. Their current-voltage characteristics reveal a step-like behavior in the vicinity of zero bias, indicating resonant tunneling of heavy-holes via three-dimensionally confined unoccupied hole states in Ge quantum dots. Assuming parabolic confinement we extract the strength of the confinement potential of quantum dots.

\end{abstract}


\maketitle

\indent The unique high frequency characteristics of resonant tunneling diodes made them suitable for use in a variety of electronics applications \cite{Brown01Paper89,Mazumder03Paper98,McMeekin01Paper96}. The progress in semiconductor technology turned materials like Si, Ge and Si$_{1-x}$Ge$_{x}$ to promising ones for the development of the semiconductor devices, based on the phenomena of quantum tunneling of charge carriers through potential barriers. There are nevertheless some natural properties of these materials restricting their applicability, for instance for optical applications due to the indirect band gap of these materials. It was shown \cite{Brunner02Review02} that Si/Ge heteroepitaxy on prepatterned substrates leads to the formation of regimented arrays of self-assembled quantum dots (QDs) with predefined ordering and geometry being an effective way to improve electronic and optical properties of Si/Ge heterostructures. Here we report on the study of confinement properties of two-terminal devices, based on the quantum tunneling through Ge QDs of definite spatial distribution.

\indent The double-barrier resonant tunneling diodes (DBRTD), investigated in this work, were processed of a Si/Ge heterostructure, grown by molecular beam epitaxy on a prepatterned \textit{p}\textminus Si (100) substrate with a resistivity of 0.02\textminus 0.08~$\Omega$. To obtain a substrate having a two-dimensional (2D) pit pattern with a periodicity of 280$~\times$~280~nm$^2$, extreme ultraviolet interference lithography (EUV-IL) at a wavelength of 13.5~nm  was used \cite{Auzelyte01Paper09,Dais02Paper08}. The EUV-IL was performed at the Swiss Light Source at the Paul Scherrer Institut and the light exposing dose was chosen to be as high as 250~mW/cm$^{2}$. The exposed pit pattern was transferred into the Si substrate by reactive ion etching down to a depth of 10 nm \cite{Dais02Paper08}.

\indent The first layer grown on the substrate is a 250~\AA\ thick boron doped \textit{p}$^{+}$\textendash Si buffer layer (\textit{p} $=$ 5 $\times$ 10$^{18}$~cm$^{-3}$). It is followed by 200~\AA\ of strained \textit{p}$^{+}$-doped Si$_{1-x}$Ge$_{x}$ with graded Ge content ($x$) ranging from $x=$ 0.05 to 0.30 and the same doping concentration as in the Si buffer layer. After this first Si$_{1-x}$Ge$_{x}$ film an undoped Si barrier with a thickness of 50~\AA\ is grown. Next, six monolayers (6 ML) of strained Ge are deposited. By growing of a Ge film with a thickness of 6 ML, we exceed the critical thickness for the realization of 2D planar deposition of Ge on Si, which is about 3 monolayers. This exceedance leads to the formation of self-assembled Ge QDs in our heterostructure. The second barrier of identical parameters as the first one clads the germanium film. This barrier is in turn overgrown by the second graded \ Si$_{1-x}$Ge$_{x}$ layer, similar to the first one. At last a 7000~\AA\ thick \textit{p}$^{+}$-doped Si cap layer follows. All the layers except that of Ge were grown at a substrate temperature $T=$ 300~\celsius. During the growth process of Ge the temperature was ramped from 300 up to 470~\celsius\ to improve the homogeneity of the dot nucleation and their spatial distribution.

\indent  Despite the low deposition temperature strain- and temperature-induced intermixing of the Ge islands with Si of the lower barrier takes place \cite{Patriarche01Paper00}, leading to the formation of Si$_{1-x}$Ge$_{x}$ alloy with a nonhomogeneous distribution of Ge content  \cite{Kirfel01Paper04}. The photoluminescence and X\textendash ray diffractometry measurement data, obtained on similarly grown heterostructures, let us assume an average Ge content in our QDs to be about 50~\%.

\begin{figure} [t]
\noindent\centering{
 \includegraphics[width=1.0\columnwidth]{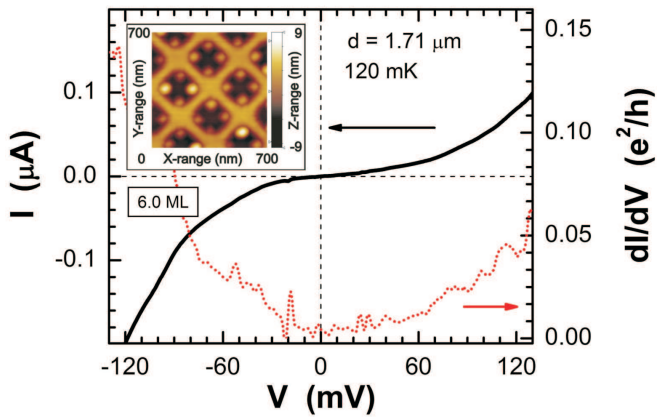}
}
\caption{IV-characteristic (solid curve) and its derivative (dotted curve) of DBRTD with a cross-section diameter d  $=$\ 1.71~$\mu$m, containing 6 monolayers of Ge quantum dots,  at $T=$ 120~mK. No magnetic field is applied. The inset shows AFM surface scan of the prepatterned test sample after the growth of 6 ML of Ge. The formation of regimented quantum dot arrays consisting of up to four quantum dots (lightly colored circles) at the corners of etched pits (dark rectangular regions) is obviously seen.}
\label{pic:FigKont02IVCbig}
\end{figure}

\indent An uncapped test sample with 6 ML of Ge (exposed to EUV light of the same dose of 250~mW/cm$^{2}$) was grown to investigate its topography. The inset of Fig.~\ref{pic:FigKont02IVCbig} shows an atomic force microscopy (AFM) image of its surface.  The interplay of strain and surface tension of the deposited Ge layer results in the formation of regimented arrays of up to four (105)-faceted 3D Ge hut clusters (further referred to simply as QDs) per prepatterned pit.

\indent A strained Si/Ge heterostructure with Ge QDs forms a type-II heterojunction. Its band alignment creates a potential well only for the holes within the Ge dots \cite{Yakimov02Paper01}.  As our heterostructure is strained, the heavy-hole (HH) band in the Si$_{1-x}$Ge$_{x}$ emitter becomes decoupled from the light-hole (LH) and split-off bands \cite{Pikus}. We assume that  due to the band splitting only HH states are occupied in the emitter at $T\sim$ 100~mK. Thus, we consider only HH take part in the charge transport.

\indent The vertical DBRTDs were processed using electron-beam lithography, wet-etching and polyimide isolation technique to support the top contacting pads \cite{Reed02Paper88,Tewordt01Paper91}. The diameter of the diodes varies between 1 and 2~$\mu$m. Transport measurements were made in a dilution refrigerator in the temperature range from 100~mK to 1~K. A homogeneous magnetic field $B$ was applied perpendicular (angle 90$^\circ$ in our notation) and parallel (0$^\circ$) to the heterointerface. The current-voltage (IV) characteristics were measured in a 2-point measurement technique. The forward bias refers to the positively biased bottom electrode (substrate) of the diode.

\indent Figure~\ref{pic:FigKont02IVCbig} shows the IV-characteristic (along with its derivative $dI/dV$) of a vertical DBRTD with the cross-section diameter of d = 1.71~$\mu$m. The measurement was carried out at $B=0$ and $T=$ 120~mK.
We do not observe, on the characteristics of our diodes with QDs, any strongly expressed resonant current peak which is typical for a DBRTD with a quantum well. The reason for the absence of a peak could lay in the complicated heterostructure interface profile due to the substrate prepatterning along with the possible dopant diffusion from the highly doped Si$_{1-x}$Ge$_{x}$ regions into the barriers due to the absence of an undoped spacer between them, which could destroy the resonant tunneling characteristics. We emphasize that our samples do not contain a quantum well but quantum dots.

\indent The clearly nonlinear IV-characteristic reveals a lot of features in its derivative. The slight asymmetry of the characteristic in forward and reverse bias is due to the difference in the substrate and top electrode doping \cite{Zaslavsky01Paper92}. In the bias range $\pm$60~mV a set of small current steps on the IV-characteristic along with the corresponding peaks in the $dI/dV$ are resolved for both polarities of applied voltage. At higher bias this peak structure washes out.

\begin{figure} [!h]
\noindent\centering{
\includegraphics[width=1.0\columnwidth]{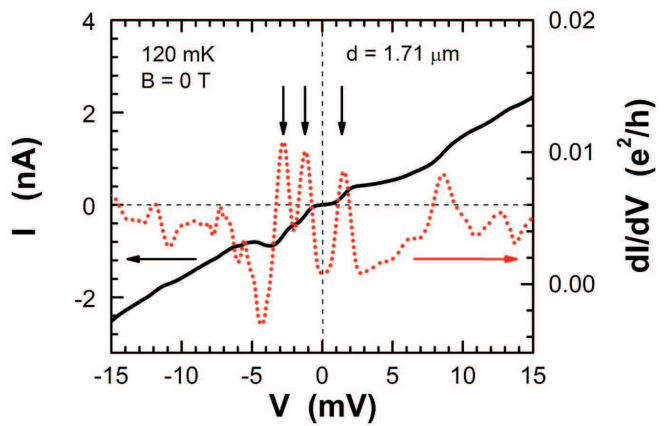}
}
\caption{IV-characteristic (solid curve) and its derivative (dotted curve) of a DBRTD (6 ML Ge), having a diameter $d=$ 1.71~$\mu$m, in the vicinity of zero bias.  $T=$~120~mK, magnetic field $B=$ 0.}
\label{pic:FigKont02IVCsmall}
\end{figure}

\indent From the temperature dependent broadening of the correspondent current step edge, we have determined the energy\textendash to\textendash bias conversion factor $\alpha$ \cite{Su02Paper98}. It turned out to be equal to 0.49~meV/mV for the three well pronounced $dI/dV$ peaks in the vicinity of zero bias, which are marked by vertical arrows in Fig.~\ref{pic:FigKont02IVCsmall}. There are other $dI/dV$ peaks that appear at higher voltages ranging from $\pm$15 to $\pm$35~mV. For these peaks we find $\alpha\approx$ 0.15~meV/mV, which is quite different compared to $\alpha$ for the peaks mentioned above. Therefore, the observed peaks can be divided  into two groups according to the correspondent values of $\alpha$. We attribute the peaks in the vicinity of zero bias to resonant tunneling of HH from the emitter electrode into the three-dimensionally confined discrete HH states within the QDs. Taking into account an average distance of 117~nm between the QDs within the arrays, we conclude that the above mentioned tunneling occurs through the individual quantum dots and not over the array as a whole.  The group of peaks at higher voltages is assigned to tunneling events through impurities located outside the Ge layer, namely, in the Si barriers.

\indent Magnetotunneling spectroscopy has proven to be a very powerful and informative method for the investigation of electron \cite{Goldman01Paper87,Leadbeater03Paper89} and hole \cite{Eaves03Paper92,LiuJ03Paper00} dynamics in DBRTDs. We have applied magnetic fields $B$ up to 12~T parallel ($B_{\parallel}$) and perpendicular ($B_{\perp}$) to the heterointerface to resolve the strength of the confinement potential and to investigate its effect on the transport properties of our DBRTD.

\begin{figure} [t]
\noindent\centering{
\includegraphics[width=0.8\columnwidth]{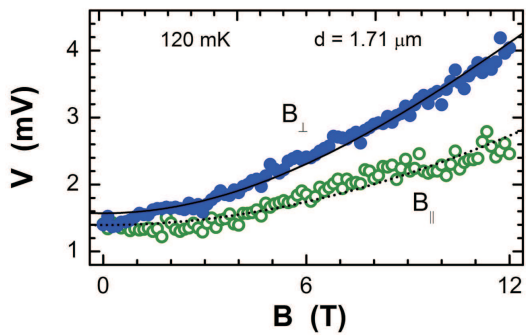}
}
\caption{Magnetic field dependence of the differential conductance peak at bias +1.4~mV. Filled circles (\textcolor[rgb]{0.00,0.00,0.50}{\Circsteel}) correspond to the 90$^\circ$ angle between \textbf{$\vec{B}$} and the plane of the heterointerface. Open circles (\textcolor[rgb]{0.00,0.40,0.29}{\Circpipe}) are the data obtained at an angle of 0$^\circ$. Solid curve and dotted curve depict the best fit achieved.}
\label{pic:FigPeakTransform}
\end{figure}

\indent Under the influence of increasing magnetic fields of both $B_{\parallel}$ and $B_{\perp}$ orientations, we observed a noticeable shift of all $dI/dV$ peaks from their initial voltage position at $B=$ 0. For the peaks, appearing in the vicinity of zero bias, the shift additionally turned out to be field orientation dependent. We concentrate our discussion on these three $dI/dV$ peaks from Fig.~\ref{pic:FigKont02IVCsmall}. Figure~\ref{pic:FigPeakTransform} shows the field dependent evolution of the $dI/dV$ peak being observed at the bias $+1.4$~mV for $B=$ 0. It shows a clear difference in the peak shift under $B_{\parallel}$ (open circles) and $B_{\perp}$ (filled circles). From these observed shifts we can gain information about the hole confinement in our QDs.

\indent For the data analysis, we have made the following assumptions and simplifications.  We neglect spin effects and valence band intermixing. We treat the HH in the QDs of our sample as parabolically confined in all three spatial directions and suppose that all three $dI/dV$ peaks in the vicinity of zero voltage correspond to tunneling over the lowest HH states. The expression to describe the ground state energy of a single particle confined by a parabolic potential $V_{0}(r) = m^{*}\omega^2_{0}r^2/2$ is given by

\begin{equation}
E(B) = E_0 +  \sqrt{\frac{1}{4}(\hbar\omega_{c})^2 + (\hbar\omega_{0})^2}
\label{eq:ConfEnergy}
\end{equation}

\noindent where $E_0$ accounts for the voltage position of the correspondent $dI/dV$ peak at $B=0$, $m^{*}$ is the effective mass, $r$ is a radial distance, $\hbar\omega_{0}$ is the confinement energy and $\omega_{c}=eB/m^{*}$ is the cyclotron frequency.

\indent As was mentioned above, we suppose the Ge fraction of Si$_{1-x}$Ge$_{x}$ alloy forming the QDs to be of the order of $x=$\ 0.5.
We neglect the effect of the strain on the HH effective mass $m_{HH}^{*}$ and for simplicity treat it also as isotropic.
Thus $m_{HH}^{*}=0.243\,m_0$ is estimated according to Ref.~\cite{Paul01Review04} from the Luttinger parameters $\gamma_1$, $\gamma_2$ determined from Ref.~\cite{Schaeffler01Paper97}.
Here $m_{0}$ is the free electron mass.
Using (1) we have performed fits on the experimental voltage position data sets for all $dI/dV$ peaks discussed in this work.
The obtained fitting curves match our experimental data quite well.
In Fig.~\ref{pic:FigPeakTransform} two best-fitting curves (solid and dotted) for the $dI/dV$ peak appearing at $+1.4$~mV ($B=$ 0) are shown.
In Ref.~\cite{Rego01Paper97} the hole states in \textit{p}-type quantum disks in magnetic field were studied theoretically.
A similar increase of the lowest-lying HH ground state energy under the influence of the rising magnetic field as observed in our experiment is reported there.

\indent The important result of the fitting procedure consists in revealing a value for the QD confinement potential. For the two fit curves in Fig.~\ref{pic:FigPeakTransform} we obtained $\hbar\omega_{0}^{\parallel}\approx$ 2.64~meV under $B_{\perp}$ and $\hbar\omega_{0}^{\perp}\approx$ 5.9~meV under $B_{\parallel}$. For the other two peaks of this group similar $\hbar\omega_{0}$ values were obtained. The observed difference in $\hbar\omega_{0}$ for two mutually perpendicular magnetic field directions is consistent with the expectation, that due to the geometrical factors the lateral confinement (revealed under the $B_{\perp}$) should be weaker compared to the confinement in the QDs growth direction.

\indent As was mentioned the $dI/dV$ peaks observed far from zero voltage originate from HH resonant tunneling via impurities located within Si barriers.
Contrary to the peaks discussed above these $dI/dV$ peaks do not show field  orientational dependence of their voltage position.
This indicates the symmetry of the confinement potential.
The fitting based on the assumption of the isotropic $m_{HH}^{*}$ revealed for both magnetic field orientations almost equal values $\hbar\omega_{0}\approx$ 21~meV.  This result supports the idea about tunneling over impurities having much smaller size than QDs and thus apparently stronger confining properties.
Calculations taking into account the possible anisotropy of $m_{HH}^{*}$\ (Ref.~\cite{Schuberth01Paper91}) reveal for these peaks $\hbar\omega_{0}^{\parallel}\approx 3\,\hbar\omega_{0}^{\perp}$ what would contradict our experimental observations.

\indent To conclude, we have investigated the electric transport properties of DBRTDs with regularly ordered self-assembled Ge quantum dots.  Steplike structures in the IV-characteristic and corresponding sharp peaklike $dI/dV$ structures are revealed.
We attribute the $dI/dV$ peaks in the vicinity of zero voltage to resonant tunneling of HH through confined discrete HH states within QDs. The peaks at higher voltage originate from the resonant tunneling via impurities located in Si barriers.
With the help of magnetotunneling spectroscopy, assuming a parabolic confinement and an isotropy of $m_{HH}^{*}$ we have obtained values for the strength of the confinement potential for impurities and for quantum dots.

\end{document}